\documentclass[11pt,a4paper]{article}
\pdfoutput=1
\usepackage{jheppub}
\usepackage[hyperref,svgnames]{xcolor}
\usepackage[latin1]{inputenc}
\usepackage{amsmath}
\usepackage{amsfonts}
\usepackage{amssymb}
\usepackage{booktabs}
\usepackage{graphicx}
\usepackage{setspace}
\usepackage{mathrsfs}
\usepackage{tikz}
\usepackage{caption}
\usepackage{subcaption}

\definecolor{hgreen}{rgb}{0,.3,0}
\definecolor{hred}{rgb}{.3,0,0}
\definecolor{hblue}{rgb}{0,0,.3}
\definecolor{LightGray}{gray}{0.95}

\numberwithin{equation}{section}

\title{Higgs Assisted Q-balls from Pseudo-Nambu-Goldstone Bosons\\
}
\author[1]{Fady Bishara}
\emailAdd{fady.bishara@physics.ox.ac.uk}

\author[1]{George Johnson}
\emailAdd{george.johnson@physics.ox.ac.uk}

\author[1]{Olivier Lennon}
\emailAdd{olivier.lennon@physics.ox.ac.uk}

\author[1]{John March-Russell}
\emailAdd{jmr@thphys.ox.ac.uk}

\affiliation[1]{Rudolf Peierls Centre for Theoretical Physics, University of Oxford, 1 Keble Road, Oxford OX1 3NP, United Kingdom}

\newcommand{\nf}{N_f}
\newcommand{\sun}[1]{{SU}(#1)}
\newcommand{\sunf}{\sun{\nf}}
\newcommand{\un}[1]{{U}(#1)}

\newcommand{\avg}[1]{\mathrm{tr} \left( #1\right)}
\newcommand{\delsig}[2]{\partial_{#1}^{#2}\Sigma}
\newcommand{\delsighc}[2]{\partial_{#1}^{#2}\Sigma^\dagger}
\newcommand{\sighc}{\Sigma^\dagger}

\newcommand{\mcD}{\mathcal{D}}

\newcommand{\mcL}{\mathcal{L}}
\newcommand{\mcO}{\mathcal{O}}

\newcommand{\mbbH}{H}
\newcommand{\mbbS}{S}

\newcommand{\vev}{v_h}

\newcommand{\eul}{\mathcal{E}_\omega}
\newcommand{\mpi}{m_\pi}

\newcommand{\lamp}{\lambda_p}
\newcommand{\lams}{\lambda_s}
\newcommand{\lamh}{\lambda_h}
\newcommand{\Eul}{energy functional }
\newcommand{\lamchihs}{\Lambda_\chi^\textsc{hs}}

\newcommand{\nh}{\eta}
\newcommand{\minim}{minimal}
\newcommand{\full}{full}
  
\abstract{Motivated by recent constructions of TeV-scale strongly-coupled
	dynamics, either associated with the Higgs sector itself as in
	pseudo-Nambu-Goldstone boson (pNGB) Higgs models or in theories of asymmetric
	dark matter, we show that stable solitonic Q-balls can be formed from light
	pion-like pNGB fields carrying a conserved global quantum number in the presence
	of the Higgs field.   We focus on the case of thick-wall Q-balls, where
	solutions satisfying all constraints are shown to exist over a range of
	parameter values.  In the limit that our approximations hold, the Q-balls are
	weakly bound and parametrically large, and the form of the interactions of the
	light physical Higgs with the Q-ball is determined by the breaking of scale
	symmetry.}

\graphicspath{{./figs/}}

\preprint{OUTP-17-10P}

\begin{document}

\tikzstyle{every picture}+=[remember picture]
\usetikzlibrary{shapes.geometric}
\usetikzlibrary{calc}
\usetikzlibrary{decorations.pathreplacing}
\usetikzlibrary{decorations.markings}
\usetikzlibrary{decorations.text}
\usetikzlibrary{patterns}
\usetikzlibrary{backgrounds}
\usetikzlibrary{positioning}
\tikzstyle arrowstyle=[scale=2]
\tikzstyle directed=[postaction={decorate,decoration={markings,
		mark=at position 0.6 with {\arrow[arrowstyle]{>}}}}]
\tikzstyle rarrow=[postaction={decorate,decoration={markings,
		mark=at position 0.999 with {\arrow[arrowstyle]{>}}}}]

\everymath{\displaystyle}

\maketitle

\section{Introduction}

Stable soliton-like solutions exist in a wide variety of quantum field theories
in four, and other, dimensions.  Broadly speaking, solitons may be characterised
as either \emph{topological} or \emph{non-topological} solitons.  Topological
solitons have their stability guaranteed by the conservation of a suitable
topological `charge' or winding number.  For example, `t~Hooft-Polyakov
monopoles in spontaneously broken non-Abelian (3+1)-dimensional gauge theories
are characterised by the second homotopy group of the vacuum manifold and the
associated winding numbers. Alternatively,  for non-topological-solitons
stability is commonly ensured by a combination of energy-conservation and a
conserved Noether charge (see, for example,~\cite{Lee:1991ax} and references
therein).  One particularly noteworthy class of such solitons are
Q-balls~\cite{Coleman:1985ki, Safian:1987pr, Lee:1988ag, Kusenko:1997ad}:
semiclassical configurations of underlying Noether-charge-carrying scalar
fields, and possibly other, additional fields too.

In most studies of Q-ball solutions, the scalar fields making up the Q-ball are
explicitly or implicitly assumed to be elementary.  For instance, Q-balls that
are absolutely stable, or metastable with cosmological lifetimes, have been
studied in supersymmetric extensions to the Standard Model where the underlying
scalar fields are combinations of the elementary Higgs, slepton, and/or squark
fields of the model~\cite{Kusenko:1997si, Kusenko:2001vu}. These solitons are
intrinsically interesting objects to study theoretically, and often have the
additional intriguing property that they can provide potential dark matter
candidates~\cite{Kusenko:1997si, Kusenko:2001vu, Graham:2015apa}, so may be of
relevance to phenomenology.

In this work we show that Q-balls can exist in theories where the charged scalar
fields that make up the Q-ball are not elementary but rather composite states,
with non-perturbative dynamics leading to a low-energy effective theory
described by light pion-like pseudo-Nambu-Goldstone bosons (pNGBs) carrying a
$\un{1}$ global quantum number.  In particular, with an eye towards future
possible applications to Beyond-the-Standard Model and dark matter physics, we
consider theories which contain a strongly-interacting hidden sector at
TeV-scales or above, and which feature a spontaneous breaking of a non-Abelian
global symmetry similar to that of the chiral symmetry breaking of QCD, but
occurring at $f\sim$~TeV energies or greater, rather than the scale
$f\sim$~100~MeV as for QCD.   When small explicit breaking of the original
global symmetry is included, the previously massless Nambu-Goldstone bosons
acquire small masses. Importantly, these, now pseudo-Nambu-Goldstone bosons, can
be much lighter than all other mass scales associated with the strongly-coupled
sector, and so we can treat their low-energy dynamics separately from all other
degrees of freedom originating from the strongly-coupled theory.

For our purposes it is also important that the pNGB fields can also naturally
carry a variety of conserved $U(1)$ `flavour' quantum numbers.\footnote{These
are independent of a possible $U(1)$ `baryon number' acting on states of the UV
theory.}  As the pNGBs are the lightest charged states transforming under these
$U(1)$'s,  and in addition the pNGBs interact via both derivative and
\emph{non-derivative} potential terms, in principle it is possible for stable
Q-balls formed out of these pNGBs to exist.\footnote{Here we of course require
that the analogue of flavour-violating weak interactions in the hidden sector
are either not present in the theory, or the interactions mediated by these
additional forces do not violate the global $U(1)$ we will be considering.  We
will discuss simple examples where this is the case in Section~\ref{structure}.}
However, in the case that the \emph{only} light fields are the pNGBs, the
leading ${\cal O}(p^2)$ effective Lagrangian describing the low-energy
interactions of the pNGBs fails to satisfy the energetic conditions necessary for a
Q-ball to be stable against decay into individual pNGBs. This situation can be
altered by inclusion of ${\cal O}(p^4)$ terms, but only at the expense of rather
large second-order coefficients~\cite{Distler:1986ta}.

Fortunately, in the situation we study in this work, the pNGBs are not the only
relevant light fields.  In general the Standard Model Higgs field is even
lighter than the hidden pNGBs and, as we show in Section~\ref{structure},
interacts with them in a particular way via a Higgs-portal interaction.  The
form of the resulting pNGB-Higgs interactions is not arbitrary, but constrained
by the breaking of scale symmetry~\cite{Voloshin:1980zf,
Voloshin:1985tc,Chivukula:1989ds}. This then leads to an interacting system of
both charge-carrying and charge-neutral scalar fields that we show in
Section~\ref{main} possesses Q-ball solutions for a range of underlying
parameter values. Specifically, in this paper we focus on the existence of
small-to-moderate-charge `thick-wall' Q-ball solutions which we find are
applicable in a charge range up to at least $Q\sim 10^4$, and sometimes $Q\sim
10^8$ depending on underlying parameters of the model -- see
Eq.~\eqref{eq:thickwallbound} and
Figs.~\ref{fig:energy}-\ref{fig:radius-charge}. For this reason we expect them
to be of greater phenomenological relevance than thin-wall Q-balls.  In this
work we solely consider the existence and properties of these thick-wall Q-ball
solutions, leaving their possible phenomenological applications to a later
study.

Before turning to the details of our particular model and the existence of Q-ball solutions, we emphasise that the underlying UV strong-coupling dynamics plays almost no role in the analysis,\footnote{The exception being the presence or otherwise of a Fermi repulsion term depending on the fermion or boson nature of the underlying matter degrees of freedom in the UV theory.} the existence and detailed properties of the Q-ball solutions depending solely on the leading-order low-energy effective Lagrangian interactions between the pNGBs themselves and with the Higgs.   We therefore expect that similar Q-ball solutions will occur in a wide range of effective field theories described by the Callan-Coleman-Wess-Zumino coset construction \cite{Coleman:1969sm,Callan:1969sn} supplemented by Higgs interactions.  In particular it would be interesting to study the possible existence of stable or metastable Q-balls in models where the Higgs doublet itself is realised as a pNGB, along with other light pNGB fields~\cite{Kaplan:1983fs,Kaplan:1983sm,Georgi:1984af,Dugan:1984hq,Agashe:2004rs,Bellazzini:2014yua}.  

Finally, it is worth mentioning that it is not strictly necessary that the $U(1)$ which
stabilises our thick-wall Q-balls is an exact global symmetry.  Small breaking by
higher-order terms suppressed by a high scale would render the Q-ball unstable
but long-lived, similar to the manner in which conventional neutron stars are still
cosmologically long-lived objects in the presence of sufficiently small
baryon number violation.  Alternatively, the stabilising $U(1)$ could even be an
unbroken gauge symmetry, since for a gauge coupling that is parametrically
smaller than the Higgs-pNGB interaction strength, our thick-wall Q-balls would be
unperturbed to leading order~\cite{Lee:1988ag}.

\section{The structure of the model}
\label{structure}

We assume that there are two sectors: the Standard Model (SM) and a hidden sector (HS). As described in the Introduction, the HS possesses a spontaneously broken  almost-exact global symmetry, which gives rise to pNGBs. The HS also possesses an unbroken global $U(1)$,  under which some of these pNGBs transform.
In this section we describe the origin of the Higgs coupling to the HS pions resulting in the Lagrangian in Eq.~(\ref{eq:lag-higgs-coupling}).

\subsection{Structure of the hidden sector}

For definiteness, we consider a HS with a QCD-like $SU(N_c)$ Yang-Mills theory
with $N_f$ flavours of HS `quarks' in the fundamental of $SU(N_c)$.  This theory
possesses an $\sunf_L\times \sunf_R$ chiral flavour symmetry which is
spontaneously broken to the diagonal subgroup $\sunf_V$.\footnote{We ignore the
fact that the symmetry group is generally $\un{\nf}_L\times \un{\nf}_R$ since
the one non-anomalous $U(1)$ from the $\un{\nf}_L\times \un{\nf}_R$, that in the
SM case corresponds to baryon number, acts trivially on the pNGBs, so it is not
of interest to us here.} Then, by Goldstone's theorem, there will be $\nf^2-1$
massless Nambu-Goldstone bosons that parameterise the coset space
$\sunf_L\times\sunf_R/\sunf_V$. Furthermore, the HS quark mass matrix, $M$,
explicitly breaks the chiral symmetry, which becomes only approximate in this
limit.  The NGBs will therefore acquire a non-zero mass, i.e., they become
pNGBs. The mass matrix $M$ also breaks $\sunf_V$ if it is not proportional to
the unit matrix: in the situation that no two HS quark masses are equal, the
surviving global symmetry acting on the pNGBs is $U(1)^{\nf-1}$, in the absence
of other interactions.

As usual, we can describe the light pNGBs transforming under the non-linearly realised $\sunf_L\times \sunf_R$ symmetry by a unitary matrix field
of unit determinant built from the $\nf^2-1$ pNGBs, $\pi^a$:
\begin{equation}
\Sigma=\exp(i \pi^a T^a / f) . \label{eq:sigma}
\end{equation}
Then, under the global vectorial symmetry, $\Sigma$ transforms as
\begin{equation}
\Sigma\to\Sigma'=V\,\Sigma\,V^\dagger,
\label{eq:sigma-trans}
\end{equation}
where $V$ is in general given by $V=\exp(-iX)$ with $X$ Hermitian and traceless. The Noether current associated to this transformation is
\begin{equation}
J^\mu=
	i\frac{f^2}{4}\avg{
		[\Sigma^\dagger,X] \partial^\mu\Sigma+[\Sigma,X]\partial^\mu\Sigma^\dagger
		},
\label{eq:noether-current}
\end{equation}
where we have assumed the usual leading order chiral Lagrangian 
\begin{equation}
\mcL = \frac{f^2}{4}\avg{\delsig{\mu}{}\delsighc{}{\mu}} - \frac{B_0 f^2}{2} \avg{M(\Sigma + \sighc - 2)}.
\label{eq:chiral-lag}
\end{equation}
For generic diagonal $M$, the transformation Eq.~\eqref{eq:sigma-trans} is a symmetry when $X$ is one of the possible $N_f-1$ diagonal matrices.  The pseudoscalar sector can both possess global symmetries, and have a non-trivial potential
given by the second term in Eq.~\eqref{eq:chiral-lag}, so it is reasonable to ask whether stable Q-balls can be present in this sector. However, to address this question (which ultimately requires some numerical analysis) we need to be more specific about the coupling of this HS to the SM, and also about the HS itself, as well as the exact form of the global $U(1)$ that we will be using.

For concreteness, suppose that the HS is very similar in form to the SM itself,
but with the analogue of $U(1)_{Y}$ \emph{ungauged}.  Thus we take the HS gauge
group to be $\sun{3}'\times\sun{2}'$ with, minimally, one `generation' of matter
fermions in the same $\sun{3}'\times\sun{2}'$ representations as the SM matter
fields.  This guarantees the anomaly freedom of the matter content with respect
to these two symmetries.  We also require the HS quarks to acquire bare masses,
so the HS must also have an $\sun{2}'$-doublet scalar state, $S$, which acquires
a vacuum expectation value (VEV), analogous to the Higgs doublet in the SM
sector. This scalar doublet is coupled via Yukawa interactions to HS chiral
quarks which acquire a mass upon the spontaneous breaking of $\sun{2}'$. The
Yukawa terms in our HS Lagrangian are
\begin{equation}
\mathcal{L}^{\text{HS}}\supset y_{ij}\overline{Q}_{L,i}Sq_{R,j}+\text{h.c.},
\label{eq:yukawa}
\end{equation}
where $Q$ ($q$) are the doublet (singlet) HS quarks and $y_{ij}$ are Yukawa couplings.  

\begin{figure}\centering
	\begin{tikzpicture}[line cap=round,
	decoration={brace,amplitude=7pt}]
		\def\dx		{1.0};	\def\ddx	{0.1};	\def\vevh	{-2.25};
		\def\vevs	{-0.6};	\def\lamx	{0.0};	\def\mpi	{-1.25};
		\def\dheavy	{0.6}
		\node  [anchor=west] at (\dx+\ddx,\vevs) {$v_s$};
		\node  [anchor=west] at (\dx+\ddx,\lamx) {$\Lambda_\chi^\text{\sc hs}$};
		\node  [anchor=west] at (\dx+\ddx,\mpi) {HS pions $\sim\mathcal{O}$(few TeV)};
		\node  [anchor=west] at (\dx+\ddx,\vevh) {$v_h$};
		\node  [anchor=south] at (0,\lamx+1) {$\vdots$};
		\draw [thick] (\dx,\vevs) -- (-\dx,\vevs);
		\draw [thick] (\dx,\lamx+0.6 ) -- (-\dx,\lamx+0.6 );
		\draw [thick] (\dx,\lamx+0.75 ) -- (-\dx,\lamx+0.75 );
		\draw [ultra thick, line cap=round] (\dx,\lamx) -- (-\dx,\lamx);
		\draw [thick] (\dx,\mpi) -- (-\dx,\mpi);
		\draw [thick] (\dx,\mpi+0.1) -- (-\dx,\mpi+0.1);
		\draw [thick] (\dx,\mpi-0.1) -- (-\dx,\mpi-0.1);
		\draw [thick] (\dx,\vevh) -- (-\dx,\vevh);
  \draw [decorate,very thick,align=left] (\dx+0.3,\lamx+\dheavy+1.1) -- (\dx+0.3,\lamx+\dheavy)
  node [midway,anchor=west,inner sep=2pt, outer sep=10pt]{heavy HS mesons\\and baryons};
	\end{tikzpicture}
	\caption{Spectrum of hidden sector states.}
	\label{fig:SpectrumHS}
\end{figure}
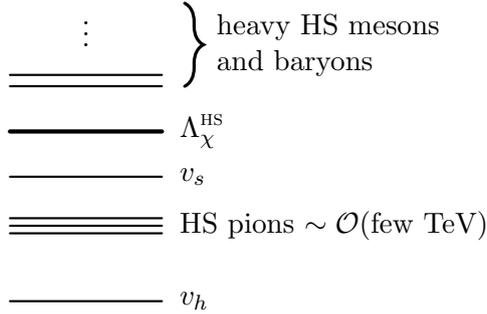

The $\sun{3}'$ is asymptotically free and confines at low energies with a
corresponding confinement scale $\lamchihs$.  We require the HS quarks of the
first generation to be light relative to $\lamchihs$ and assume that any additional
generations beyond the first are heavy.\footnote{This setup has obvious
similarities with Mirror World~\cite{Kobzarev:1966qya, Foot:1991bp, Foot:2014mia} and Twin Higgs~\cite{Chacko:2005pe,Chacko:2005vw,Chacko:2005un} scenarios, and in particular
the Fraternal Twin Higgs
models~\cite{Craig:2015pha,Garcia:2015loa,Garcia:2015toa,Craig:2015xla,Farina:2016ndq}, although in our case we are taking the HS $\sun{3}'$ dynamical scale $\lamchihs\gtrsim1\,\mathrm{TeV}$ rather than the few GeV appropriate for the Twin Higgs models.} The light HS quarks will then hadronise into a massive but light triplet of HS pions as a consequence of chiral symmetry breaking -- see Fig.~\ref{fig:SpectrumHS} for a schematic of the spectrum.

To ensure that the HS pions are absolutely stable in the presence of $\sun{2}'$ interactions, the HS `leptons' -- minimally one generation -- must have masses above the HS pion masses.

\subsection{Coupling the two sectors}

The leading interaction between the two sectors is due to a Higgs-portal interaction.  Specifically,
the scalar potential for the SM Higgs and the HS doublet is given by
\begin{equation}
V(\mbbH,\mbbS) = -\mu_h^2 \mbbH^\dagger \mbbH + \lamh(\mbbH^\dagger \mbbH)^2 - \mu_s^2 \mbbS^\dagger\mbbS + \lams(\mbbS^\dagger\mbbS)^2
+ \lamp(\mbbH^\dagger\mbbH) (\mbbS^\dagger\mbbS)\,.
\label{eq:scalar-pot}
\end{equation}
This potential induces spontaneous symmetry breaking in both sectors. We write the VEV of $S$ as $\left<S\right>= v_{s}/\sqrt{2}$ and the VEV of $H$ as $\left<H\right> = v_h / \sqrt{2}$, with $v_h = 246\,\mathrm{GeV}$ the electroweak VEV.

We introduced a portal term in the above Lagrangian with a coupling $\lamp$.  This is a marginal operator which can arise from integrating out heavier degrees of freedom and is allowed by the symmetries of the two sectors.
The portal coupling results in the mixing of the SM and HS Higgs gauge eigenstates $h'$ and $s'$ into the mass eigenstates $h$ and $s$ (see Appendix~\ref{app:scalar-masses} for details).
Working in the small mixing angle limit, $\theta\ll 1$, $s'$ can be written in terms of the mass eigenstates as
\begin{equation}
s' \approx 
s -\theta h \approx s -\frac{\lamp v_h}{2\lams v_s} h .
\label{eq:sprime-to-h-trans}
\end{equation}

Furthermore, when the HS pions are heavier than the lightest mass eigenstate $h$, the form of the couplings of $h$ to the HS pions is fully determined by the breaking of scale symmetry in the HS theory. In particular, following the work of Voloshin and Zakharov~\cite{Voloshin:1980zf,Voloshin:1985tc}, later explicated by Chivukula et al.~\cite{Chivukula:1989ds},
we may first write down the effective chiral Lagrangian for the interactions with the HS gauge eigenstate $s'$, which is given at leading order by
\begin{equation}
\mcL = \left(1+\frac{4n_h}{3\beta_0}\frac{s'}{v_s}\right)
	\frac{f^2}{4}\avg{\delsig{\mu}{}\delsighc{}{\mu}}
	+ \left(1+\left[1+\frac{2n_h}{\beta_0}\right]\frac{s'}{v_s}\right)
		\frac{B_0 f^2}{2} \avg{M(\Sigma + \sighc - 2)}.
\label{eq:lag-sprime-coupling}
\end{equation}
Here, $n_h$ is the number of heavy flavours -- i.e., the number of quarks $q$ with $m_q>\Lambda_\chi^\text{\sc hs}$ -- and $\beta_0$ is the one-loop beta function in the HS, given for general $\sun{N_c}$ with $n_\ell$ light flavours by
\begin{eqnarray}
\beta_0 = \frac{1}{3}\left(11C_A-4T_F n_\ell\right),
\label{eq:beta0}
\end{eqnarray}
where $C_A=N_c$ and $T_F=1/2$ sets the normalisation of the generators.
%To gain intuition for the additional terms in the chiral Lagrangian, note that the terms proportional to the number of heavy quarks come from integrating them out, with each heavy quark contributing the same term (for details of the numerical coefficients, see~\cite{Chivukula:1989ds}). The final additional term comes from the Yukawa terms for the light quarks, now bound in the HS pions.
The additional terms in~Eq.~\eqref{eq:lag-sprime-coupling} that couple $s'$ to the HS pions originate either from integrating out heavy quarks (terms proportional to $n_h$) or via the Yukawa terms in Eq.~\eqref{eq:yukawa} -- for details of the numerical coefficients, see~\cite{Chivukula:1989ds}.
Finally, the Lagrangian in Eq.~(\ref{eq:lag-sprime-coupling}) can be rewritten in favour of $h/v_s$ using the relation in Eq.~(\ref{eq:sprime-to-h-trans}) as
\begin{equation}
\mcL = 
\left(1-\theta\frac{2\nh}{3}\frac{h}{v_s}\right)\frac{f^2}{4}\avg{\delsig{\mu}{}\delsighc{}{\mu}}
+ \left(1-\theta\left(1+\nh\right)\frac{h}{v_s}\right)
\frac{B_0 f^2}{2}\avg{M(\Sigma + \sighc - 2)},
\label{eq:lag-higgs-coupling}
\end{equation}
where we have defined $\nh\equiv2n_h/\beta_0$ (and neglected interactions with the scalar $s$ on the grounds that it is much heavier than the other scalar states).
We will show in the following section that the field theory defined by the Lagrangian Eq.~(\ref{eq:lag-higgs-coupling}) admits
thick-wall Q-ball solutions.

\section{Higgs assisted thick-wall Q-balls}
\label{main}

We will show in this section that thick-wall Q-balls formed from the light scalars of the theory (the HS pions and the SM Higgs) can exist. We will present two cases: an analytic example with no heavy quarks in the HS, and a numerical example with arbitrarily many heavy quarks. One can ask whether thick-wall Q-balls exist in the chiral Lagrangian alone, i.e., with no coupling to the Higgs field. This is not possible (see Appendix~\ref{app:ThickWallPureChPT} for details).

\subsection{Constructing and minimising the \Eul}

The thick-wall Q-ball limit corresponds to the limit in which the field values at the centre of the Q-ball are small such that terms quartic (and higher) in the fields can be neglected~\cite{Kusenko:1997ad}. Expanding $\Sigma$ in the Lagrangian given in Eq.~\eqref{eq:lag-higgs-coupling}, using Eq.~\eqref{eq:sigma}, and including terms involving the Higgs alone gives, to cubic order,\footnote{Here we have written the Higgs cubic coupling in terms of the electroweak VEV $\vev$ and a dimensionless parameter $\lambda$, which is a function of the parameters of Eq.~\eqref{eq:scalar-pot}. See Appendix~\ref{app:scalar-masses} for details.}
\begin{equation}
\begin{split}
\mcL =&
\left(1-\theta\frac{2\nh}{3}\frac{h}{v_s}\right)
\left(\partial_\mu\pi^+\partial^\mu\pi^- + \frac{1}{2}\partial_\mu\pi^0\partial^\mu\pi^0\right)\\
	&- m_\pi^2\left(1-\theta\left(1+\nh\right)\frac{h}{v_s}\right)
\left(\pi^+\pi^- + \frac{1}{2}\pi^0\pi^0\right)
	+ \frac{1}{2}\partial_\mu h \partial^\mu h - \frac{1}{2}m_h^2 h^2 -\lambda \vev h^3.
	\label{eq:cubiclagrangian}
\end{split}
\end{equation}
This Lagrangian is invariant under a global $\un{1}$ which we can take without loss of generality to act as $\pi^\pm\to e^{\pm i\alpha}\pi^\pm$, $\pi^0 \to \pi^0$ (the labels on the pions thus refer to their charge under this $\un{1}$ symmetry, not to their electromagnetic charge). The Noether current associated with this symmetry is
\begin{equation}
J_\mu = i\left(1-\theta\frac{2\nh}{3}\frac{h}{v_s}\right)\pi^+\overset{\leftrightarrow}{\partial}_\mu\pi^-.
\end{equation}
The Hamiltonian density is given by
\begin{equation}
\label{eq:HamDensity}
\begin{split}
\mathcal{H} = &\left(1-\theta\frac{2\nh}{3}\frac{h}{v_s}\right)
\left(
	\dot{\pi}^+\dot{\pi}^-+\frac{1}{2}\dot{\pi}^0\dot{\pi}^0
	+ \vec\nabla\pi^+\cdot\vec\nabla\pi^-
	+ \frac{1}{2}\vec\nabla\pi^0\cdot\vec\nabla\pi^0
\right)\\
	&+ \frac{1}{2}\dot{h}^2+\frac{1}{2}\vec\nabla h\cdot\vec\nabla h
	+ U(\vec{\pi},h),
\end{split}
\end{equation}
where the potential $U(\vec{\pi},h)$ is
\begin{equation}
U(\vec{\pi},h) \equiv m_\pi^2\left(1-\theta\left(1+\nh\right)\frac{h}{v_s}\right)
\left(\pi^+\pi^- + \frac{1}{2}\pi^0\pi^0\right) 
	+ \frac{1}{2}m_h^2 h^2 + \lambda \vev h^3.
\end{equation}

We want to minimise the total energy for a fixed Noether charge $Q>0$. This is most
easily accomplished using the method of Lagrange multipliers~\cite{Kusenko:1997zq}. To do this, we define an \Eul $\eul$ with Lagrange multiplier $\omega$:
\begin{equation}
\eul = E+\omega\left(Q-\int \text{d}^3x \; J^0 \right) ,
\label{eq:lagrangemultiplier}
\end{equation}
where $E$ is the integral of the Hamiltonian density, given in Eq.~\eqref{eq:HamDensity}. Inserting the above expressions for the Noether current and the Hamiltonian density, we obtain
\begin{align}
\label{eq:euler}
\eul = &\int \text{d}^3x&&\kern-1em\left[\left(1-\theta\frac{2\nh}{3}\frac{h}{v_s}\right)
	\left(\left|\dot{\pi}^+ - i\omega\pi^+\right|^2 
	+ \frac{1}{2}\dot{\pi}^0\dot{\pi}^0\right)
	+ \frac{1}{2}\dot{h}^2 \right.\nonumber\\
	& &&\kern-1em+\left.\left(1-\theta\frac{2\nh}{3}\frac{h}{v_s}\right)
	\left(\vec\nabla{\pi}^+\cdot\vec\nabla{\pi}^-
	+ \frac{1}{2}\vec\nabla\pi^0\cdot\vec\nabla\pi^0 \right)
	+ \frac{1}{2}\vec\nabla h\cdot\vec\nabla h
	+ \widehat{U}(\vec{\pi},h)\right]\\
	&+ \omega Q,&&\nonumber
\end{align}
where
\begin{equation}
\widehat{U}(\vec{\pi},h) = U(\vec{\pi},h) - \left(1-\theta\frac{2\nh}{3}\frac{h}{v_s}\right)\omega^2\pi^+\pi^-.
\label{eq:hatpotential}
\end{equation}

The only explicit time dependence of $\eul$ has been isolated in the first line of Eq.~(\ref{eq:euler}). This integral is positive semidefinite, and to minimise its contribution to the energy the fields must have the following time dependence: 
\begin{equation}
\pi^\pm(x,t) = e^{\pm i\omega t}\pi^\pm(x),\quad \pi^0(x,t) = \pi^0(x),\quad h(x,t) = h(x).
\end{equation}

Our problem now involves four real degrees of freedom: $\pi^\pm(x)$, $\pi^0(x)$, and $h(x)$. To proceed, we assume that the spatial profile of each of the fields has, up to normalisations, the same form:
\begin{equation}
\pi^\pm(x)=\pi(x),\qquad \pi^0(x)=\beta\pi(x),\qquad h(x)=\alpha\pi(x)\,,
\end{equation}
where we allow for $\alpha$ and/or $\beta$ to be zero.
This ansatz is sufficient for the purpose of demonstrating the existence of Q-balls; in reality, the spatial profiles of the fields might differ, but this extra freedom in the minimisation process can only further lower the Q-ball energy. With these proportionality relations, we can write $\eul$ solely in terms of the field $\pi(x)$. In addition to the gradient-squared and field-squared terms, we also have the cross-term
\begin{equation}
-\theta\frac{2\nh}{3}\frac{\alpha}{v_s}\left(1+\frac{1}{2}\beta^2\right)\pi(\vec\nabla\pi)^2.
\end{equation}
This term can be dropped to leading order in a self-consistent approximation scheme for the Q-ball solution. This is because it is suppressed relative to the $(\vec\nabla\pi)^2$ term by the mixing angle and $\langle\pi\rangle/v_s$, where $\langle\pi\rangle$ is the maximum value of the pion VEV inside the Q-ball, and to the $\pi^3$ term in Eq.~(\ref{eq:hatpotential}) by spatial gradients, which we will \textit{a posteriori} check to be small. This term is also exactly absent when $\nh\propto n_h=0$.

It now remains to minimise the \Eul with respect to the function $\pi(x)$ and the three variables $\alpha$, $\beta$, and $\omega$.
To do this, it is useful to redefine the fields and the coordinates in Eq.~(\ref{eq:euler}) in order to isolate them in a dimensionless integral (see Appendix~\ref{app:field-redefs} for details). After these redefinitions, the \Eul becomes
\begin{equation}
\frac{\eul}{Q \mpi} =
\frac{S_\psi}{Q \alpha^2}
\frac{
	\left(1+\dfrac{1}{2}\beta^2+\dfrac{1}{2}\dfrac{m_h^2}{\mpi^2}\alpha^2-\Omega^2\right)^{3/2}
	\left(1+\dfrac{1}{2}\beta^2+\dfrac{1}{2}\alpha^2\right)^{3/2}
}
{
\left(\dfrac{\mpi}{v_s}\theta(1+\nh)\left[1+\dfrac{1}{2}\beta^2-\dfrac{2\nh}{3(1+\nh)}\Omega^2\right]
	-\dfrac{\lambda\vev}{\mpi}\alpha^2\right)^2
}+\Omega,
\label{eq:dim'less-energy-per-Q}
\end{equation}
where $\Omega\equiv\omega/\mpi$ and $S_\psi$ is given by
\begin{equation}\label{eq:s_psi}
S_\psi = \int \mathrm{d}^3\xi\left(\vec\nabla_\xi \psi\cdot\vec\nabla_\xi \psi +\psi^2-\psi^3\right),
\end{equation}
where $\xi$ and $\psi$ are the spatial coordinate and field in dimensionless units, defined in Eq.~\eqref{eq:redefinitions}. This has the same form as the bounce action for an analogous Euclidean tunnelling problem in three dimensions, and so we can make use of previous results on this subject~\cite{Coleman:1977py, Callan:1977pt, Coleman:1977th}. In particular, the integral is minimised when the field is spherically symmetric, and thus we expect all Q-ball solutions to be spherically symmetric. The value of Eq.~\eqref{eq:s_psi} when minimised is approximately $38.8$~\cite{Linde:1981zj}. 

A global minimum with $\eul/Q\mpi<1$ corresponds to a classically stable Q-ball solution. After minimisation with respect to $\omega$, which enforces the fixed-charge constraint, and $\alpha$ and $\beta$, the Q-ball has a mass $M_Q = \eul$.
The radius, $R_Q$, of the Q-ball is $\sim1$ in terms of the dimensionless coordinate $\xi$. Translated into the parameters of the model, it is given by
\begin{equation}
R_Q^{-1}\sim m_\pi\frac{\left(1+\dfrac{1}{2}\beta^2+\dfrac{1}{2}\dfrac{m_h^2}{\mpi^2}\alpha^2-\Omega^2\right)^{1/2}}{\left(1+\dfrac{1}{2}\beta^2+\dfrac{1}{2}\alpha^2\right)^{1/2}}.
\label{eq:radius}
\end{equation}

We will minimise $\eul$ in two cases: first, we will analytically study the case
that there are no additional quarks with masses above the chiral symmetry
breaking scale in the HS, and with the Higgs acting as a massless mediator; second, we will numerically study the case that there are arbitrarily many heavy quarks in the HS, allowing the Higgs mass and self-coupling to be non-zero. 

\begin{figure}
\centering
\begin{subfigure}{0.49\textwidth}
  \centering
  \includegraphics[width=\linewidth]{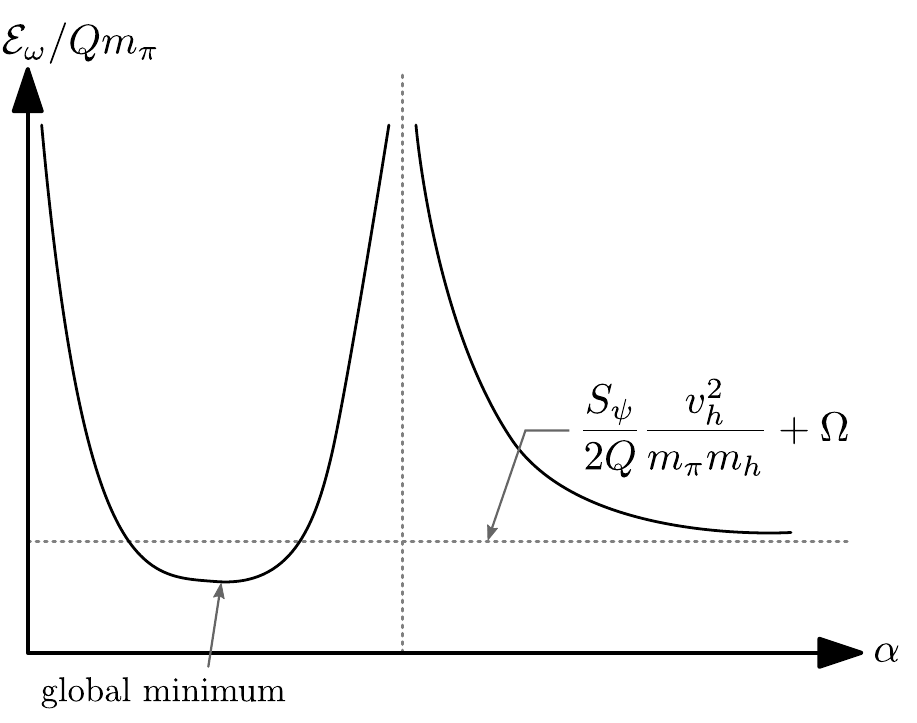}
  \label{fig:sub1}
\end{subfigure}
\begin{subfigure}{0.49\textwidth}
  \centering
  \includegraphics[width=\linewidth]{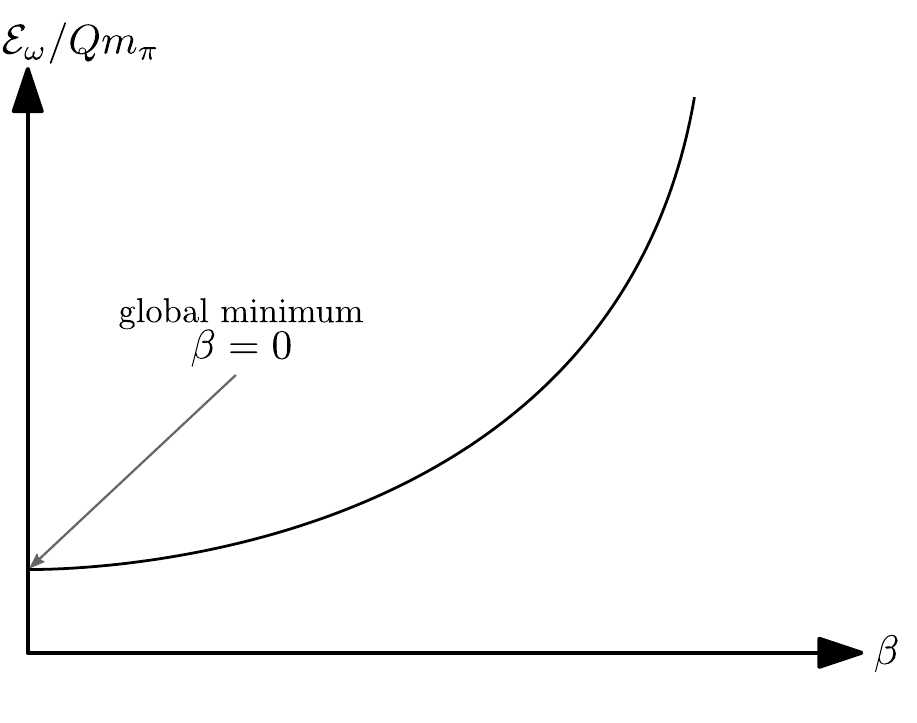}
  \label{fig:sub2}
\end{subfigure}
\caption{The schematic behaviour of the dimensionless \Eul
		$\eul/Q m_\pi$ in Eq.~(\ref{eq:dim'less-energy-per-Q}) as a function of $\alpha$ and $\beta$, for typical choices of parameters. It can be seen that the formation of a Higgs VEV inside the Q-ball is energetically favoured, whilst the formation of a neutral pion VEV is not.}
\label{fig:alphabeta}
\end{figure}

The qualitative dependence of the \Eul on $\alpha$ and $\beta$ is shown in Fig.~\ref{fig:alphabeta}, for typical parameter choices.
Physically we expect that, 
for $m_h^2/m_\pi^2\ll 1$, $\beta$ will be zero for the following reason. The neutral pion has no cubic interactions with the charged pions, unlike the Higgs, and thus no direct way to lower the energy of the Q-ball. It does, however, have a cubic interaction with the Higgs, which will acquire a VEV at the centre of the Q-ball along with the charged pions, and this cubic interaction may favour the neutral pion acquiring a VEV of its own. However, since this interaction is quadratic in the neutral pion, the Higgs VEV in the Q-ball must be sufficiently large that this term dominates the neutral pion mass term. We hence expect that for pions much heavier than the Higgs, the neutral pion will have a VEV of precisely zero. We will see that this is so in both the analytic and the numerical analysis of the subsequent two sections.

\subsection{An analytic example: no heavy quarks}
\label{sec:analytic-example}

In order to determine the conditions for the existence of Q-balls in this theory, as well as the nature of the Q-balls should they exist, we must minimise the \Eul given in Eq.~\eqref{eq:dim'less-energy-per-Q} with respect to $\Omega$, $\alpha$, and $\beta$. This is not possible to do analytically in the general case: minimising with respect to $\Omega$ requires finding the roots of a sixth-order polynomial. The barrier to analyticity comes from the term proportional to $\Omega$ in the denominator of $\eul$. Thus, to gain an analytic understanding of the Q-ball, we assume that there are no heavy quarks in the HS: this sets $\nh=0$ and therefore removes the problematic term. To make the results more straightforward and illuminating, we will also take the Higgs mass and cubic self-coupling to zero; it is possible to analytically study the system without this assumption, but at the expense of making the results more opaque. This assumption is valid provided the pion mass terms and the cubic coupling of the pions to the Higgs dominate the aforementioned terms in the Lagrangian; that is,
\begin{equation}
\frac{1}{2}\frac{m_{h}^{2}}{\mpi^{2}}\alpha^{2}\ll1 \quad \mathrm{and} \quad \frac{\lambda \vev v_s}{\theta \mpi^{2}}\alpha^{2}\ll1.
\end{equation}
We will leave a more general discussion of this type of hidden sector until Section~\ref{numerics}, where we relax this assumption and the assumption that $n_h=0$ with a numerical minimisation of the Q-ball energy, scanning over the parameters of the model.

Setting $\nh=0$ and $m_h = \lambda = 0$ in Eq.~\eqref{eq:dim'less-energy-per-Q}, we first minimise with respect to $\alpha$ to obtain $\alpha^2 = 4 + 2\beta^2$. Substituting this back into the energy functional, we observe that, for $\Omega^2 > 0$, the expression is a strictly increasing function of $\beta$, and hence is minimised when $\beta = 0$ (as argued in the previous section). Thus
\begin{equation}
\alpha = 2 \quad \mathrm{and} \quad \beta = 0.
\end{equation}
The energy of the Q-ball is minimised when the VEV of the neutral pion inside the Q-ball is zero, whilst the VEV of the Higgs is double that of the charged pions.

With these substitutions, we have an \Eul of the same form, as a function of $\Omega$, as that given in~\cite{Kusenko:1997ad}. We can therefore translate those results across to our case. The \Eul is minimised with respect to $\Omega$ if
\begin{equation}
\epsilon\equiv\frac{4}{9\sqrt{3}S_\psi}\frac{Q\theta^2\mpi^2}{v_s^2}=\Omega\left(1-\Omega^{2}\right)^{1/2},
\end{equation}
which has a solution for $\Omega$ provided $0<\epsilon<1/2$. The expression for $\Omega$ at the minimum is
\begin{equation}
\Omega=\left(\frac{1+\sqrt{1-4\epsilon^{2}}}{2}\right)^{1/2}.
\end{equation}
Substituting this back into the \Eul and expanding in $\epsilon$ yields
\begin{equation}
\label{eq:AnalyticExampleMass}
\frac{M_{Q}}{Q\mpi}=1-\frac{1}{6}\epsilon^{2}-\mathcal{O}(\epsilon^{4}),
\end{equation}
where $M_Q$ is the energy of the Q-ball. The expression on the right-hand side is clearly less than unity for $\epsilon>0$. Thus, this solution is (classically) stable for $Q>0$.\footnote{If the Higgs mass is appreciable compared to that of the pions, there is a lower bound on the charge $Q$ due to the fact that the Higgs provides an unfavourable contribution to the mass-to-charge ratio of the Q-ball.  Note also that the charge needs to be sufficiently large that quantum fluctuations are under control and the semiclassical approximation is valid. Here we take this to imply $Q\gtrsim 10$.}

From Eq.~\eqref{eq:radius} we find that the radius of the Q-ball is given by
\begin{equation}
R_Q^{-1}\sim\frac{\epsilon\mpi}{\sqrt{3}}\left(1+\frac{1}{2}\epsilon^{2}+\mathcal{O}\left(\epsilon^{4}\right)\right).
\label{eq:analytic-radius}
\end{equation}
This characteristic (inverse) length scale is proportional to the small parameter $\epsilon$, thus justifying our earlier assertion that spatial derivatives are suppressed in the thick-wall case.

Finally, the maximal value of the charged pion VEV occurs in the centre of the Q-ball and takes the value
\begin{equation}
\langle\pi(0)\rangle\sim\left(1-\Omega^{2}\right)\frac{v_s}{2\theta}\sim\left(2\times10^{-5}\right)Q^{2}\theta^{3}\left(\frac{\mpi}{v_s}\right)^{3}\mpi.
\label{eq:PionVev}
\end{equation}

This solution is subject to the following theoretical constraints. Firstly, we require the charge to be sufficiently small that the thick-wall analysis is valid. Secondly, we must check that the Q-ball number density is not so large that the Fermi degeneracy pressure due to the quarks which constitute the pions becomes important.

\subsubsection*{Thick-wall validity}The thick-wall analysis is only valid in the low charge regime. This is represented by the condition that $\epsilon<1/2$, which can be rearranged to give
\begin{equation}
Q\lesssim76\left(\frac{v_s}{\theta \mpi}\right)^{2}.
\label{eq:thickwallbound}
\end{equation}
We have assumed that the quartic terms in the \Eul are small compared to the quadratic and cubic terms, which are approximately equal in size in the centre of the Q-ball. There are two types of quartic term we need to consider: the $\pi^{4}$ term and the $h^{4}$ term.\footnote{In principle, there is also an $h^{2}\pi^{2}$ term, but this arises due to a dimension-six operator suppressed by an independent mass scale. This scale can naturally be much larger than $\lamchihs$, thus decoupling this quartic interaction.} Demanding that the Higgs quartic is indeed negligible when the pion VEV is given by its maximum value, Eq.~\eqref{eq:PionVev}, yields
\begin{equation}
Q\ll 150\frac{v_s}{\theta\mpi\sqrt{\lambda}}.
\end{equation}
Demanding that the pion quartics are negligible likewise gives the constraint
\begin{equation}
Q\ll 430 \frac{v_s f}{\theta m_\pi^2}.
\end{equation}
Note that these constraints merely place limits on the validity of the thick-wall analysis, not on the existence of a Q-ball of any description. If these constraints are strongly violated, then stable Q-balls are best described using the thin-wall analysis~\cite{Coleman:1985ki, Kusenko:1997zq}.
We will return to the issue of existence and properties of thin-wall Q-balls in this class of hidden sector models in future work. In the intermediate charge region, we expect that stable Q-balls will still exist, though these will be of neither thick- nor thin-wall type.

\subsubsection*{Fermi degeneracy pressure}
The final important consideration arises due to the fact that the scalars from which these Q-balls are built are in fact composites of fermions, the HS quarks. If the density of pions in the Q-ball is too high, Fermi degeneracy pressure due to these quarks can become significant. In this case, we expect that the radius of the Q-ball will increase to counteract this pressure and reduce the contribution to the Q-ball energy from the filled Fermi sphere. Nevertheless, we can put a conservative upper bound on the charge of the Q-ball by demanding that, for the Q-ball radius as calculated above, such energy contributions are lower than the binding energy.

In the non-relativistic limit, the average additional energy contributed to the Q-ball per constituent fermion is
\begin{equation}E = \frac{3}{5}E_F = \frac{3}{10 m_f} (3 \pi^2 n)^{2/3},\end{equation}
where $m_f$ is the fermion mass and $n$ its number density. We will demand that
\begin{equation} 2QE < Q m_\pi - M_Q .\end{equation}
This leads to 
\begin{equation}
Q \lesssim 0.1 \left(\frac{m_f}{m_\pi} \right)^{3/2} .
\label{eq:fermi}
\end{equation}
We hence see that Fermi degeneracy pressure can be quite significant. Given that the pions are pseudo-Nambu-Goldstone bosons of an approximate spontaneously-broken chiral flavour symmetry, we expect them to be relatively light compared to the other scales in the theory. In particular, the appropriate masses of the constituent (dressed) quarks should be of order the chiral symmetry breaking scale, $\lamchihs$. This is undetermined and can in principle be arbitrarily high; as such, we will not worry further about this constraint.

One might wonder whether the chiral symmetry breaking scale, if sufficiently high, might give rise to unnaturally large corrections to the Higgs mass through pion loops. The cubic Higgs-pion coupling gives rise to corrections merely logarithmic in $\lamchihs/\mpi$, however, and therefore naturalness is not a concern in this case.

\subsection{A numerical example: arbitrarily many heavy quarks}  \label{numerics}

The task of analytically minimising the energy functional, Eq.~(\ref{eq:euler}), is intractable in the
general case, but can be done numerically. In this section we present the results of a numerical minimisation of the \Eul with respect to $\alpha$, $\beta$, and $\Omega$ for various choices of $n_h$, scanning over the parameters in Eq.~(\ref{eq:euler}). Across the entirety of parameter space we find that the \Eul is minimised when $\beta = 0$. This is in line with the heuristic argument presented in Section~\ref{sec:analytic-example} that the neutral pion should not acquire a VEV inside the Q-ball.

The results are almost entirely independent of the number of heavy quarks. This is perhaps to be expected, since the number of heavy quarks enters only through a small modification to the denominator of Eq.~\eqref{eq:dim'less-energy-per-Q}. Consequently, we have chosen to use $n_h = 4$ as an illustrative example of the full numerical analysis; the most important differences between the analytic and numerical results arise from neglecting the Higgs mass and cubic coupling in the former case. We therefore also present a numerical analysis where we take $n_h = 0$, $m_h /\mpi \to 0$ and $\lambda \to 0$; this `minimal' case is meant as a cross-check against the analytic example discussed in Section~\ref{sec:analytic-example}.

\begin{table}[t]\centering
	\begin{tabular}{c c c}\toprule[1pt]
		Parameter & Range & Distribution\\\midrule
		$Q$		&	$[1,10^8]$	&	log-uniform\\
		$v_s$ [TeV]	&	$[1,10]$	&	log-uniform\\
		$m_\pi$ [TeV]	&	$[0.5,2v_s]$	&	log-uniform\\
		$\theta$	&	$[10^{-4},0.1]$	&	log-uniform\\
		$\lambda$	&	$[10^{-6}, 10^{-1}]$	&	log-uniform\\\bottomrule[1pt]
	\end{tabular}
	\caption{Scan parameters and their ranges. Log-uniform means uniformly distributed on a logarithmic scale.}
	\label{tab:scan-params}
\end{table}

The parameters were randomly sampled uniformly on a logarithmic scale. They are listed, along with their lower and upper bounds used for the scan, in Table~\ref{tab:scan-params}. A set of randomly chosen parameters was rejected if it resulted in an energetically unfavourable solution -- i.e., if Eq.~(\ref{eq:euler}) had no minimum such that $\eul/Q m_\pi < 1$. The Higgs cubic coupling $\lambda$ was treated as an independent parameter since it is poorly constrained by LHC Higgs measurements~\cite{ATL-PHYS-PUB-2014-019,ATL-PHYS-PUB-2015-046}.

\begin{figure}[t]\centering
	\begin{minipage}{\columnwidth}
		\includegraphics[width=0.49\columnwidth]{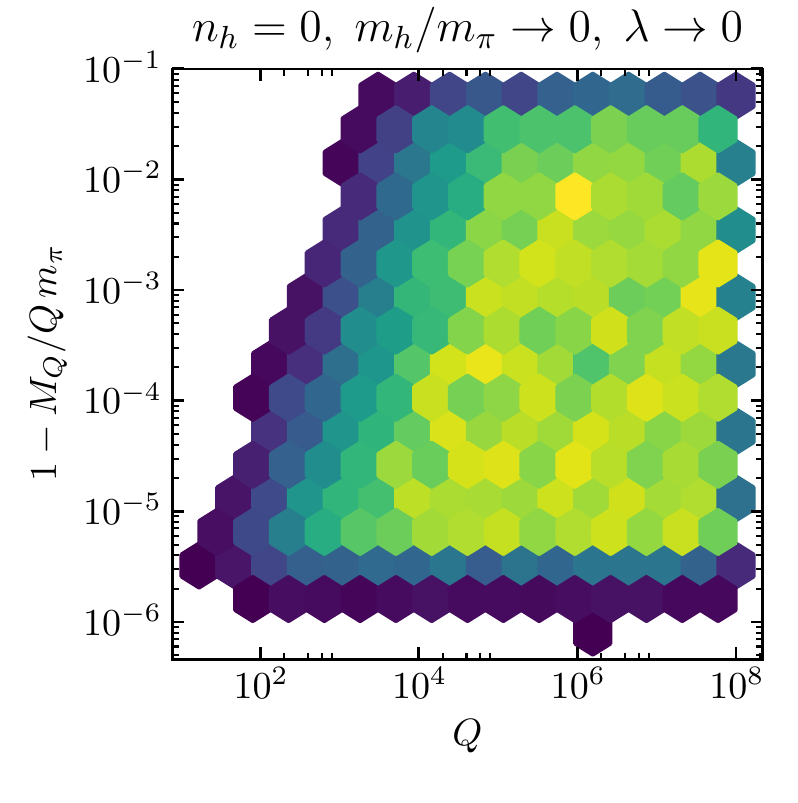}
		\includegraphics[width=0.49\columnwidth]{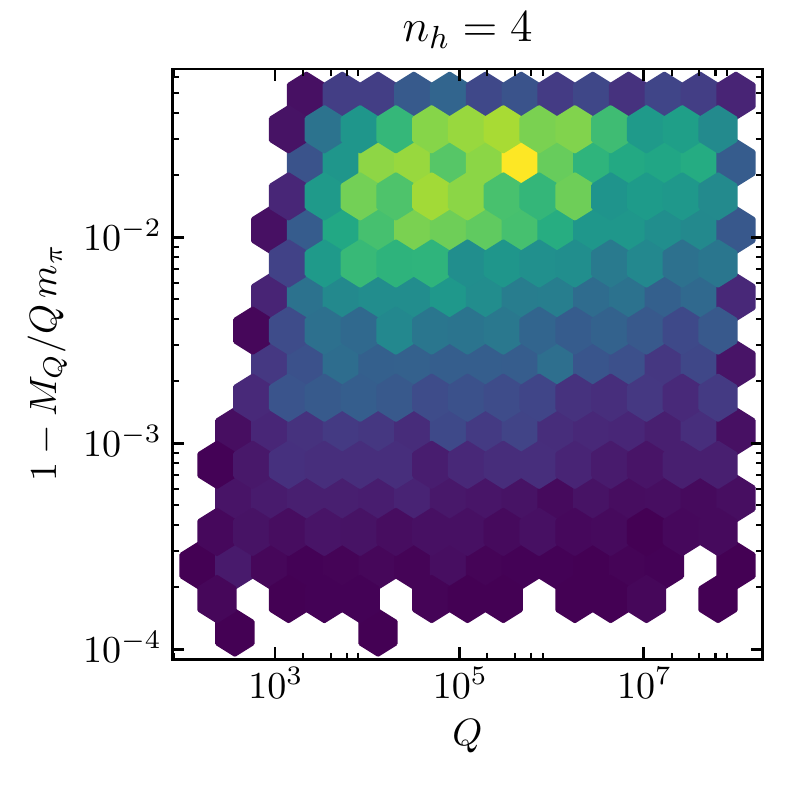}
	\end{minipage}
	\caption{
	Number of Q-ball solutions with different fractional binding energies for different choices of Q-ball charge. The shade of a given cell corresponds to the number of solutions in that cell; a lighter (more yellow) shade corresponds to more solutions. The left and right panels show the result of the scan for the \minim\ case and the \full\ case respectively.
		}
	\label{fig:energy}
\end{figure}

In the following figures, the solutions are clustered in cells and the cell brightness is directly proportional to the number of solutions it contains; the lighter (more yellow) the cell, the
larger the number of solutions contained in it. In each figure, the left (right) panel
shows the results for the minimal (\full) case.

Figure~\ref{fig:energy} shows the result of the scan for the
fractional binding energy, $1-M_Q/Qm_\pi$, versus the total Q-ball charge. The figure shows that thick wall Q-balls exist for a wide range of charges (indeed, across the entire range of charges scanned over), with (for small charges) there being a preference for larger binding energy the larger the charge. This is consistent with the expression Eq.~\eqref{eq:AnalyticExampleMass} in the analytic example. When the Higgs mass is appreciable, there is some preference for larger binding energy, across a range of charges. This can be attributed to the fact that the Higgs mass results in an unfavourable contribution to the Q-ball energy, and so favourable contributions from the other terms in the \Eul are required to be larger to offset this. The typical scale of the binding energy is thus increased.

\begin{figure}[t]\centering
	\begin{minipage}{\columnwidth}
		\includegraphics[width=0.49\columnwidth]{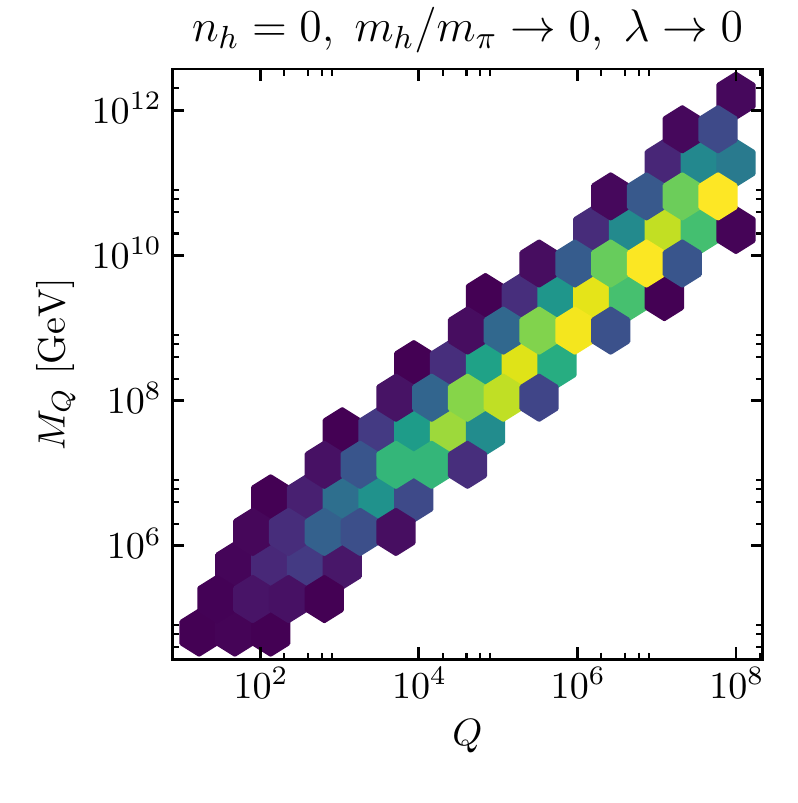}
		\includegraphics[width=0.49\columnwidth]{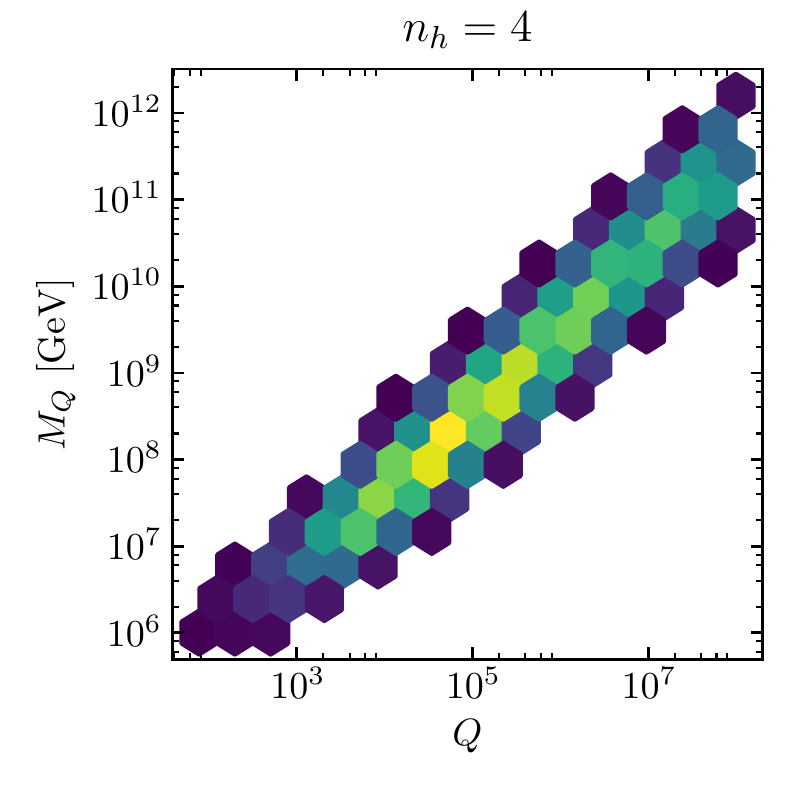}
	\end{minipage}
	\caption{The distribution of Q-ball solutions in the mass-charge plane. See caption of Fig.~\ref{fig:energy} for more details.
		}
	\label{fig:mass-charge}
\end{figure}
\begin{figure}[t]\centering
	\begin{minipage}{\columnwidth}
		\includegraphics[width=0.49\columnwidth]{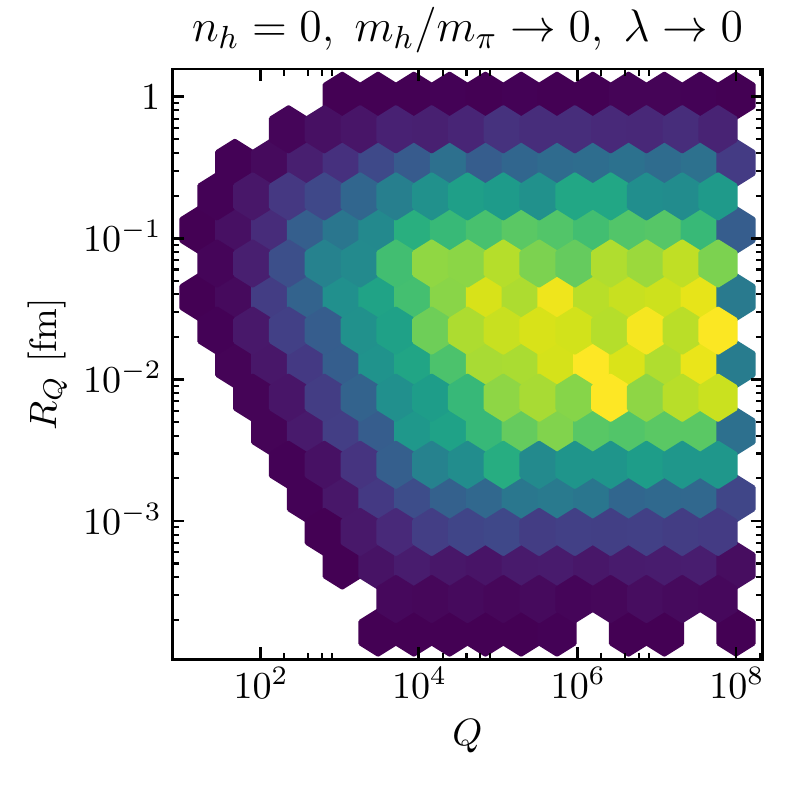}
		\includegraphics[width=0.49\columnwidth]{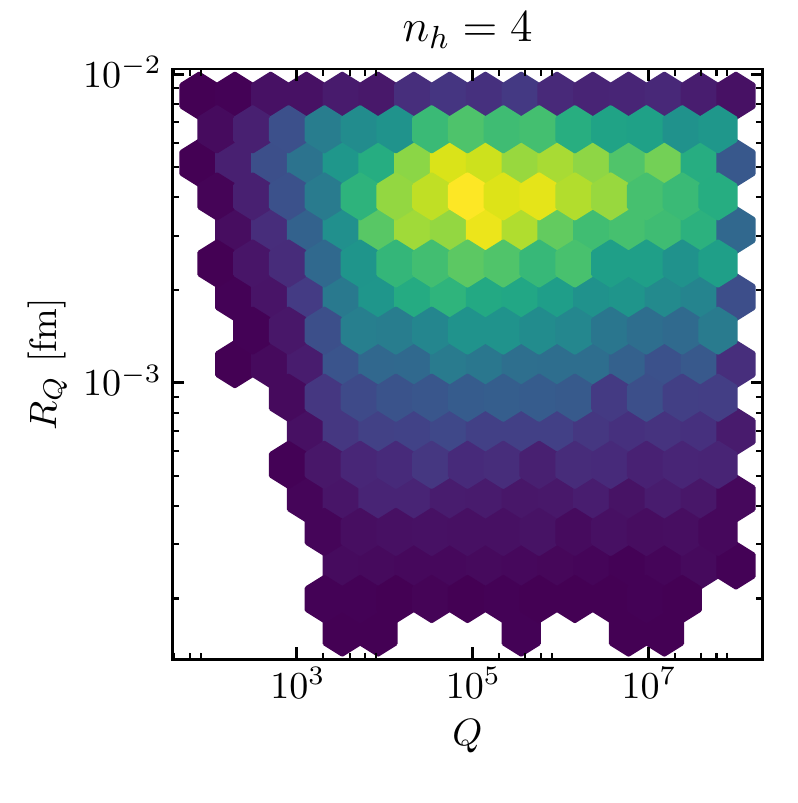}
	\end{minipage}
	\caption{The distribution of Q-ball solutions in the radius-charge plane. See caption of Fig.~\ref{fig:energy} for more details.}
	\label{fig:radius-charge}
\end{figure}

Figures~\ref{fig:mass-charge}~and~\ref{fig:radius-charge} show the behaviour of the physical Q-ball parameters, namely its mass and radius, with respect to the charge of the Q-ball. In Fig.~\ref{fig:mass-charge} there is a strong linear correlation between the mass and charge of the Q-ball in both the minimal case and the \full\ case. This is consistent with expression Eq.~\eqref{eq:AnalyticExampleMass} in the analytic example, which predicts a linear relation between the mass and charge, to leading order.

Figure~\ref{fig:radius-charge} shows that, for a given charge, there are Q-ball solutions with radii ranging from around $10^{-3}\,\text{fm}$ to around $1\,\text{fm}$ in the minimal case. The radius (for small charges) tends to be larger on average for smaller $Q$; this is consistent with the expression Eq.~\eqref{eq:analytic-radius} in the analytic example. We also note that the radius is bounded above by about $10^{-2}\,\text{fm}$ in the \full\ case, when the Higgs mass is accounted for. This effect can be traced back to Eq.~\eqref{eq:radius}, with $m_h$ acting to reduce the radius of the Q-ball. Indeed, if we take the limit $\Omega \to 1$, then whilst the Q-ball gets arbitrarily large in the minimal case, its radius is bounded above by $\sim m_h / m_\pi$ in the \full\ case. Physically we expect a lighter Higgs to yield a longer range attractive force, in turn stabilising bigger Q-balls. 

\begin{figure}[t]\centering
	\begin{minipage}{\columnwidth}
		\includegraphics[width=0.49\columnwidth]{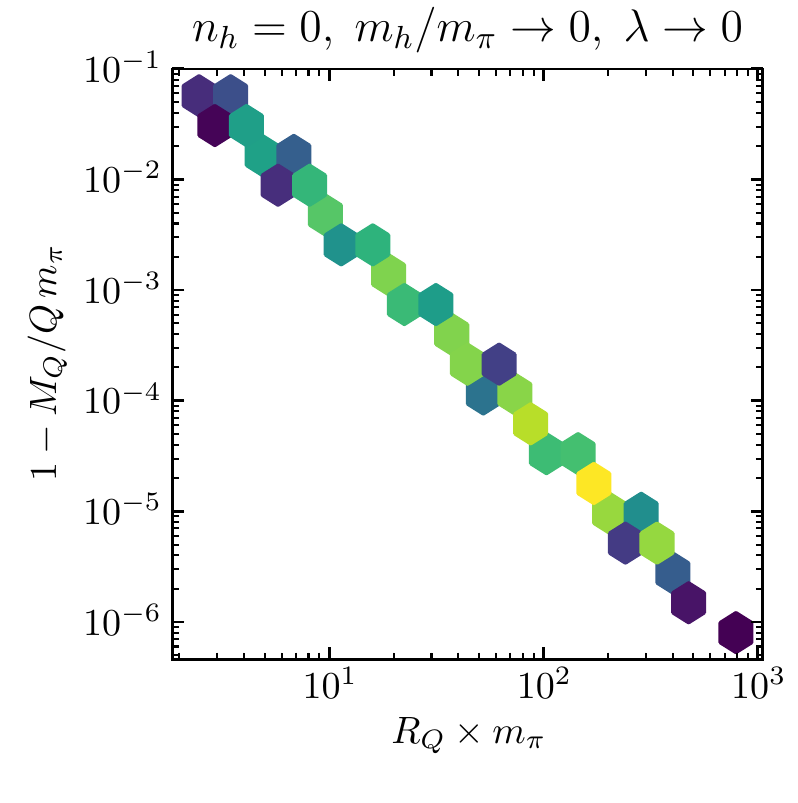}
		\includegraphics[width=0.49\columnwidth]{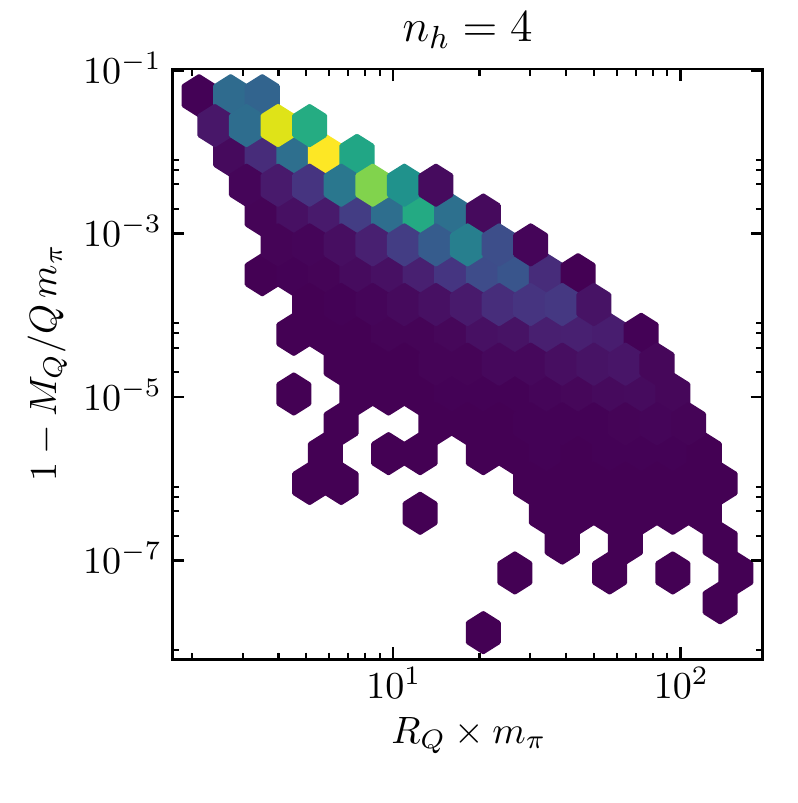}
	\end{minipage}
	\caption{
The distribution of Q-ball solutions with different fractional binding energies and radii (in units of the pion mass). See caption of Fig.~\ref{fig:energy} for more details.
		}
	\label{fig:binding-radius}
\end{figure}

Figure~\ref{fig:binding-radius} shows the relationship between the Q-ball fractional binding energy and radius, in units of the pion mass. In the minimal case there is an exact relation between these two quantities; note that Eq.~\eqref{eq:AnalyticExampleMass} and Eq.~\eqref{eq:analytic-radius} are both functions solely of $\epsilon$. To leading order this relation is linear with gradient $-2$. In the full case there is no such fixed relation, but nevertheless the binding energy is bounded above for a given radius, with there being a preference for binding energies close to this bound.

\section{Summary and conclusions}

In this work we have demonstrated, by both analytic and numerical methods, the
existence of Q-ball solutions in an interacting, hidden sector pNGB-Higgs boson
system. The specific class of low-energy effective Lagrangians we study,
Eq.~\eqref{eq:lag-higgs-coupling}, are simple generalisations of the usual
chiral Lagrangian to hidden sector QCD-like strong dynamics, supplemented by
Higgs-portal-mediated interactions with the (lighter) physical Higgs boson. We
find, in the small-to-moderate charge range ($10\lesssim Q\lesssim10^{4-8}$) we
study, that thick-wall Q-ball solutions exist. These Q-balls are relatively
weakly bound, Eq.~\eqref{eq:AnalyticExampleMass} and Fig.~\ref{fig:energy}, and have
size parametrically large compared to the inverse pNGB mass,
Eq.~\eqref{eq:analytic-radius} and Fig.~\ref{fig:radius-charge}. The range of
Q-ball properties that we find numerically are illustrated in
Figs.~\ref{fig:energy}-\ref{fig:binding-radius}. We emphasise that we have shown
that Q-balls can exist in theories where the global charge-carrying states are
composite, rather than elementary, scalars.

Such Q-ball solutions may be relevant to dark matter properties in a variety of Beyond-the-Standard-Model theories, in particular those of asymmetric dark matter and pNGB-Higgs theories. To assess whether this is the case requires a dedicated study of Q-ball production dynamics in the early Universe. Naively, there is no analogue of a decay of an Affleck-Dine condensate~\cite{Kusenko:1997si, Enqvist:1997si} as applies in supersymmetric Q-ball models of dark matter. Thus we are left with solitosynthesis and aggregation build up along the lines of~\cite{Frieman:1989bx, Griest:1989bq, Hardy:2014mqa} as the likely dominant mechanism, though the details are different. Although we expect that we never reach the thin-wall limit, it would also be interesting to study the existence and properties of thin-wall Q-ball solutions in pNGB-Higgs systems.

\section*{Acknowledgments}

FB, GJ, and OL are supported by the Science and Technology
Facilities Council (STFC).

\appendix

\section{Scalar masses and mixing angle}
\label{app:scalar-masses}

The scalar potential Eq.~\eqref{eq:scalar-pot} is generically minimised when both $|H|$ and $|S|$ acquire non-zero VEVs, which we write as $v_h/\sqrt{2}$ and $v_s/\sqrt{2}$ respectively. Expanding around these VEVs and diagonalising the resulting quadratic terms in the potential gives the masses $m_h$ and $m_s$ of the light and heavy scalar mass eigenstates of the theory. These can be read-off directly from~\cite{Barger:2008jx}. We have
\begin{align}
m_h^2 &= \lamh v_h^2+\lams v_s^2 - \sqrt{\mcD}\label{eq:higgs-mass-squared}\\
m_s^2 &= \lamh v_h^2+\lams v_s^2 + \sqrt{\mcD},
\end{align}
where
\begin{equation}
\mcD = \left(\lamh v_h^2 - \lams v_s^2   \right)^2 + \lamp^2v_h^2v_s^2.
\end{equation}
The two scalar mass eigenstates $h$ and $s$ are related to the gauge eigenstates $h'$ and $s'$ by the rotation that enacts the aforementioned diagonalisation. That is to say,
\begin{equation}
\begin{pmatrix}
h' & s'
\end{pmatrix}
\cdot M^2 \cdot
\begin{pmatrix}
h'\\s'
\end{pmatrix}
= 
\begin{pmatrix}
h&s
\end{pmatrix}
\cdot \widehat{M}^2 \cdot
\begin{pmatrix}
h\\s
\end{pmatrix}\quad\text{with}\quad
\begin{pmatrix}
h\\s
\end{pmatrix}
= 
\begin{pmatrix}
\cos\theta & -\sin\theta\\
\sin\theta & \cos\theta
\end{pmatrix}
\begin{pmatrix}
h'\\s'
\end{pmatrix},
\end{equation}
where $\widehat{M} = \mathrm{diag}(m_h, m_s)$. We identify the lightest scalar mass eigenstate, $h$, as the SM Higgs. It is the coupling of the pions to this particle that is of most interest to us, on account that it can mediate a long-range attractive force between the pions by virtue of its relative lightness. Given that it is the HS gauge eigenstate $s'$ which couples to the pions, it is necessary to find an expression for the mixing angle $\theta$.
We have
\begin{equation}
\tan(2\theta) = \frac{\lamp z}{\lams z^2-\lamh },
\label{eq:tan2phi}
\end{equation}
where $z$ is defined as ratio of the VEVs, $z\equiv v_s/v_h$. For $z$ large, assuming $\lamh$ and $\lams$ are comparable in size, we can write the mixing angle in Eq.~(\ref{eq:tan2phi}) in terms of the small parameter $\zeta\equiv z^{-1}$,
\begin{equation}
\tan(2\theta) = \frac{\lamp \zeta}{\lams - \lamh \zeta^2}
= \frac{\lamp}{\lams}\zeta+\mcO(\zeta^3).
\end{equation}
In this limit, the small angle approximation for $\theta$ is also valid and we find that
\begin{equation}
\theta\approx \frac{\lamp v_h}{2\lams v_s}.
\end{equation}
Finally, can write the SM Higgs cubic coupling $\lambda\vev$ which appears in Eq.~\eqref{eq:cubiclagrangian} in terms of the couplings in the scalar potential. We have
\begin{equation}
\lambda\vev = \lamh v_h \cos^3 \theta - \lams v_s \sin^3 \theta  +\frac{\lamp}{2}\left(v_h \cos \theta\sin^2 \theta - v_s \sin \theta \cos^2 \theta\right),
\end{equation}
and so
\begin{equation}
\lambda  \approx \lambda_h - \frac{\lambda_p^2}{4 \lambda_s}.
\label{eq:lambda}
\end{equation}

\section{Absence of thick-wall Q-balls in the pure chiral Lagrangian}
\label{app:ThickWallPureChPT}

Here we show that thick-wall Q-balls cannot exist within the leading order $\sun{2}$ chiral Lagrangian. To do this, we need to show that the functional Eq.~\eqref{eq:lagrangemultiplier}, using the Lagrangian Eq.~\eqref{eq:chiral-lag} and current Eq.~\eqref{eq:noether-current}, has no minima for $Q \neq 0$. Just as in Eq.~\eqref{eq:euler}, we can write the functional as a sum of time-dependent and time-independent pieces:
\begin{align}
\mathcal{E}_{\omega}=&\,\frac{f^{2}}{4}\int\text{d}^{3}x \,\avg{\left|\dot{\Sigma}-i\omega[\Sigma,X]\right|^{2}+\vec\nabla\Sigma^\dagger \cdot \vec{\nabla}\Sigma-2B_{0}M\left(\Sigma+\Sigma^{\dagger}-2\right)-\omega^{2}[\Sigma,X][X,\Sigma^{\dagger}]}\nonumber\\
&+\omega Q.
\end{align}
The first term in the integral contains all of the time dependence, and is positive semidefinite. Thus the functional is minimised by choosing
\begin{equation}
\Sigma(x,t)=\exp(-i\omega X t)\Sigma_{0}(x)\exp(i\omega X t),
\end{equation}
where $\Sigma_{0}(x)$ is an $\sun{2}$ matrix which depends only upon spatial coordinates. Substituting this into the above functional and choosing $X=\sigma_{3}/2$, we find
\begin{align}
\mathcal{E}_{\omega}=&\int\text{d}^{3}x&&\kern-1em\left[\left(\frac{1}{2}\vec\nabla\pi^{0}\cdot\vec\nabla\pi^{0}+\vec\nabla\pi^{+}\cdot\vec\nabla\pi^{-}\right)\left(1 - \frac{1}{3f^2}\left(\pi^0 \pi^0 + 2 \pi^+ \pi^-\right)\right) \right.\nonumber\\& &&\kern-1em+\frac{1}{6f^{2}} \left(\pi^0\vec\nabla\pi^{0}+\pi^+\vec\nabla\pi^{-} + \pi^- \vec{\nabla}\pi^+\right)^{2}+\frac{1}{2}\mpi^{2}\pi^{0} \pi^{0}+(\mpi^{2}-\omega^{2})\pi^{+}\pi^{-}\nonumber\\& &&\kern-1em\left.-\frac{\mpi^{2}}{24f^{2}}(\pi^{0})^{4} -\frac{1}{6f^{2}}(\mpi^{2}-2\omega^{2})(\pi^{0} \pi^0 )(\pi^{+}\pi^{-})-\frac{1}{6f^{2}}\left(\mpi^{2}-4\omega^{2}\right)(\pi^{+}\pi^{-})^{2}\right]\nonumber\\
&+\omega Q,&&
\end{align}
where we have defined $\mpi^{2}\equiv B_{0}\text{tr}M$, and expanded $\Sigma_0$ to quartic order in the $\pi$ fields (using Eq.~\eqref{eq:sigma}) on account that there are no cubic terms in the chiral Lagrangian. As is usual in the thick-wall analysis, we ignore higher-order terms, which will stabilise the potential. This integral is exactly that describing tunnelling through a quartic potential barrier in three dimensions~\cite{Coleman:1977py, Callan:1977pt, Coleman:1977th}, with the potential having the schematic form $U(\pi)\sim m^{2}\pi^{2}-\lambda\pi^{4}$. The solutions to the associated bounce equation are spherically symmetric.

The quartic terms containing derivatives are suppressed relative to the kinetic terms by a factor of $f^{2}$ and to the other quartic terms by spatial gradients, which are small. We will hence ignore these terms.

Notice that in the limit $\omega\rightarrow \mpi$, i.e., the thick-wall or small-field limit, the last two quartic terms have positive coefficients. In order for a potential barrier to exist (and, therefore, a bounce solution to exist), we require that the overall contribution of all three quartic terms be negative. Consequently, the VEV of the neutral pion in the centre of the Q-ball must be large relative to that of the charged pions, but since this will contribute a large amount of mass to the Q-ball without contributing to its charge, we might expect that no stable Q-balls exist.

To see this quantitatively, we relate the profiles of the pion fields as before: $\pi^{0}(x)=\beta\pi(x)$ and $\pi^{\pm}(x)=\pi(x)$. We thus find that
\begin{equation}
\begin{split}
\mathcal{E}_{\omega}=&\int\text{d}^{3}x\left[\left(1+\frac{1}{2}\beta^{2}\right)\vec\nabla\pi\cdot\vec\nabla\pi+\mpi^{2}\left(1+\frac{1}{2}\beta^{2}-\Omega^{2}\right)\pi^{2}-\lambda(\Omega,\beta)\pi^{4}\right]\\
&+\Omega Q\mpi,
\end{split}
\end{equation}
where $\Omega\equiv\omega/m_\pi$, and
\begin{equation}
\lambda(\Omega,\beta)\equiv\frac{\mpi^{2}}{6f^{2}}\left[\frac{\beta^{4}}{4}-\beta^2(2\Omega^{2}-1)-(4\Omega^{2}-1)\right]
\end{equation}
is the quartic coupling, which must be positive. Choosing
\begin{equation}
\begin{split}
\xi_{i}&=\mpi\left(1+\frac{1}{2}\beta^{2}\right)^{-1/2}\left(1+\frac{1}{2}\beta^{2}-\Omega^{2}\right)^{1/2}\,x_i ,\\
\psi&=\frac{1}{\mpi}\left(1+\frac{1}{2}\beta^{2}-\Omega^{2}\right)^{-1/2}\lambda(\Omega,\beta)^{1/2}\,\pi,
\end{split}
\end{equation}
we may transform $\mathcal{E}_{\omega}$ to
\begin{equation}
\mathcal{E}_{\omega}=\mpi\frac{\left(1+\beta^{2}/2\right)^{3/2}\left(1+\beta^{2}/2-\Omega^{2}\right)^{1/2}}{\lambda(\Omega,\beta)}S_{\psi,4}+\Omega Q\mpi ,
\end{equation}
where
\begin{equation}
S_{\psi,4}=\int\text{d}^{3}\xi\left(\vec\nabla_{\xi}\psi\cdot\vec\nabla_{\xi}\psi+\psi^{2}-\psi^{4}\right)
\end{equation}
is a positive, dimensionless number~\cite{Linde:1981zj}, whose precise value will not concern us in the following.

Now $\mathcal{E}_{\omega}$ must be minimised with respect to $\Omega$ and $\beta$. Minimising with respect to $\Omega$ yields
\begin{equation}
0 = \frac{\partial\mathcal{E}_{\omega}}{\partial\Omega}=\frac{7m_\pi^3 S_{\psi,4}\Omega}{6 f^2 \lambda(\Omega,\beta)^{2}}\frac{(1 + \beta^2/2)^{5/2}}{(1 + \beta^2/2 - \Omega^2)^{1/2}}\left(1 + \frac{1}{2}\beta^2 - \frac{4}{7} \Omega^2\right)+Q\mpi .
\end{equation}
This is positive semidefinite for $\Omega \in [0,1]$, vanishing only when $Q=0$ and $\Omega=0$. As such, there is no Q-ball solution.

\section{Field redefinitions}
\label{app:field-redefs}

The following rescalings of the spatial coordinates $x_i$ and the field $\pi$ are necessary to remove all parameters of the theory from inside the integral in Eq.~\eqref{eq:euler}:
\begin{equation}
\begin{split}
\xi_{i}&=m_\pi\frac{\left(1+\dfrac{1}{2}\beta^2+\dfrac{1}{2}\dfrac{m_h^2}{\mpi^2}\alpha^2-\Omega^2\right)^{1/2}}{
	\left(1+\dfrac{1}{2}\beta^2+\dfrac{1}{2}\alpha^2\right)^{1/2}}\,x_i,\\
\psi&=\alpha\frac{
	\dfrac{\mpi}{v_s}\theta(1+\nh)\left(1+\dfrac{1}{2}\beta^2-\dfrac{2\nh}{3(1+\nh)}\Omega^2\right)
	-\dfrac{\lambda\vev}{\mpi}\alpha^2
	}
	{
		m_\pi\left(1+\dfrac{1}{2}\beta^2+\dfrac{1}{2}\dfrac{m_h^2}{\mpi^2}\alpha^2-\Omega^2\right)
	}\,\pi.
\end{split}
\label{eq:redefinitions}
\end{equation}
These redefinitions allow us to minimise the resulting dimensionless integral, via the calculus of variations, in a manner independent of the parameters of the theory.

\bibliographystyle{JHEP}
\bibliography{qb-chi-lag}

\providecommand{\href}[2]{#2}\begingroup\raggedright\begin{thebibliography}{10}

\bibitem{Lee:1991ax}
T.~D. Lee and Y.~Pang, \emph{{Nontopological solitons}},
  \href{https://doi.org/10.1016/0370-1573(92)90064-7}{\emph{Phys. Rept.}
  {\bfseries 221} (1992) 251--350}.

\bibitem{Coleman:1985ki}
S.~R. Coleman, \emph{{Q Balls}},
  \href{https://doi.org/10.1016/0550-3213(85)90286-X}{\emph{Nucl. Phys.}
  {\bfseries B262} (1985) 263}. [Erratum:
  \href{https://doi.org/10.1016/0550-3213(86)90520-1}{Nucl. Phys. B269, 744
  (1986)}].

\bibitem{Safian:1987pr}
A.~M. Safian, S.~R. Coleman and M.~Axenides, \emph{{Some non-Abelian Q-balls}},
  \href{https://doi.org/10.1016/0550-3213(88)90315-X}{\emph{Nucl. Phys.}
  {\bfseries B297} (1988) 498--514}.

\bibitem{Lee:1988ag}
K.-M. Lee, J.~A. Stein-Schabes, R.~Watkins and L.~M. Widrow, \emph{{Gauged Q
  Balls}}, \href{https://doi.org/10.1103/PhysRevD.39.1665}{\emph{Phys. Rev.}
  {\bfseries D39} (1989) 1665}.

\bibitem{Kusenko:1997ad}
A.~Kusenko, \emph{{Small Q balls}},
  \href{https://doi.org/10.1016/S0370-2693(97)00582-0}{\emph{Phys. Lett.}
  {\bfseries B404} (1997) 285},
  [\href{https://arxiv.org/abs/hep-th/9704073}{{\ttfamily hep-th/9704073}}].

\bibitem{Kusenko:1997si}
A.~Kusenko and M.~E. Shaposhnikov, \emph{{Supersymmetric Q balls as dark
  matter}}, \href{https://doi.org/10.1016/S0370-2693(97)01375-0}{\emph{Phys.
  Lett.} {\bfseries B418} (1998) 46--54},
  [\href{https://arxiv.org/abs/hep-ph/9709492}{{\ttfamily hep-ph/9709492}}].

\bibitem{Kusenko:2001vu}
A.~Kusenko and P.~J. Steinhardt, \emph{{Q ball candidates for selfinteracting
  dark matter}},
  \href{https://doi.org/10.1103/PhysRevLett.87.141301}{\emph{Phys. Rev. Lett.}
  {\bfseries 87} (2001) 141301},
  [\href{https://arxiv.org/abs/astro-ph/0106008}{{\ttfamily
  astro-ph/0106008}}].

\bibitem{Graham:2015apa}
P.~W. Graham, S.~Rajendran and J.~Varela, \emph{{Dark Matter Triggers of
  Supernovae}}, \href{https://doi.org/10.1103/PhysRevD.92.063007}{\emph{Phys.
  Rev.} {\bfseries D92} (2015) 063007},
  [\href{https://arxiv.org/abs/1505.04444}{{\ttfamily 1505.04444}}].

\bibitem{Distler:1986ta}
J.~Distler, B.~R. Hill and D.~Spector, \emph{{$K$ Balls in the Chiral
  Lagrangian}}, \href{https://doi.org/10.1016/0370-2693(86)91080-4}{\emph{Phys.
  Lett.} {\bfseries B182} (1986) 71--74}.

\bibitem{Voloshin:1980zf}
M.~B. Voloshin and V.~I. Zakharov, \emph{{Measuring QCD Anomalies in Hadronic
  Transitions Between Onium States}},
  \href{https://doi.org/10.1103/PhysRevLett.45.688}{\emph{Phys. Rev. Lett.}
  {\bfseries 45} (1980) 688}.

\bibitem{Voloshin:1985tc}
M.~B. Voloshin, \emph{{Once Again About the Role of Gluonic Mechanism in
  Interaction of Light Higgs Boson with Hadrons}}, {\emph{Sov. J. Nucl. Phys.}
  {\bfseries 44} (1986) 478}. [Yad. Fiz.44,738(1986)].

\bibitem{Chivukula:1989ds}
R.~S. Chivukula, A.~G. Cohen, H.~Georgi, B.~Grinstein and A.~V. Manohar,
  \emph{{Higgs decay into goldstone bosons}},
  \href{https://doi.org/10.1016/0003-4916(89)90119-X}{\emph{Annals Phys.}
  {\bfseries 192} (1989) 93--103}.

\bibitem{Coleman:1969sm}
S.~R. Coleman, J.~Wess and B.~Zumino, \emph{{Structure of phenomenological
  Lagrangians. 1.}},
  \href{https://doi.org/10.1103/PhysRev.177.2239}{\emph{Phys. Rev.} {\bfseries
  177} (1969) 2239--2247}.

\bibitem{Callan:1969sn}
C.~G. Callan, Jr., S.~R. Coleman, J.~Wess and B.~Zumino, \emph{{Structure of
  phenomenological Lagrangians. 2.}},
  \href{https://doi.org/10.1103/PhysRev.177.2247}{\emph{Phys. Rev.} {\bfseries
  177} (1969) 2247--2250}.

\bibitem{Kaplan:1983fs}
D.~B. Kaplan and H.~Georgi, \emph{{SU(2)$\times$U(1) Breaking by Vacuum
  Misalignment}},
  \href{https://doi.org/10.1016/0370-2693(84)91177-8}{\emph{Phys. Lett.}
  {\bfseries 136B} (1984) 183--186}.

\bibitem{Kaplan:1983sm}
D.~B. Kaplan, H.~Georgi and S.~Dimopoulos, \emph{{Composite Higgs Scalars}},
  \href{https://doi.org/10.1016/0370-2693(84)91178-X}{\emph{Phys. Lett.}
  {\bfseries 136B} (1984) 187--190}.

\bibitem{Georgi:1984af}
H.~Georgi and D.~B. Kaplan, \emph{{Composite Higgs and Custodial SU(2)}},
  \href{https://doi.org/10.1016/0370-2693(84)90341-1}{\emph{Phys. Lett.}
  {\bfseries 145B} (1984) 216--220}.

\bibitem{Dugan:1984hq}
M.~J. Dugan, H.~Georgi and D.~B. Kaplan, \emph{{Anatomy of a Composite Higgs
  Model}}, \href{https://doi.org/10.1016/0550-3213(85)90221-4}{\emph{Nucl.
  Phys.} {\bfseries B254} (1985) 299--326}.

\bibitem{Agashe:2004rs}
K.~Agashe, R.~Contino and A.~Pomarol, \emph{{The Minimal composite Higgs
  model}}, \href{https://doi.org/10.1016/j.nuclphysb.2005.04.035}{\emph{Nucl.
  Phys.} {\bfseries B719} (2005) 165--187},
  [\href{https://arxiv.org/abs/hep-ph/0412089}{{\ttfamily hep-ph/0412089}}].

\bibitem{Bellazzini:2014yua}
B.~Bellazzini, C.~Cs\'{a}ki and J.~Serra, \emph{{Composite Higgses}},
  \href{https://doi.org/10.1140/epjc/s10052-014-2766-x}{\emph{Eur. Phys. J.}
  {\bfseries C74} (2014) 2766},
  [\href{https://arxiv.org/abs/1401.2457}{{\ttfamily 1401.2457}}].

\bibitem{Kobzarev:1966qya}
I.~{\relax Yu}. Kobzarev, L.~B. Okun and I.~{\relax Ya}. Pomeranchuk, \emph{{On
  the possibility of experimental observation of mirror particles}},
  {\emph{Sov. J. Nucl. Phys.} {\bfseries 3} (1966) 837--841}. [Yad.
  Fiz.3,1154(1966)].

\bibitem{Foot:1991bp}
R.~Foot, H.~Lew and R.~R. Volkas, \emph{{A Model with fundamental improper
  space-time symmetries}},
  \href{https://doi.org/10.1016/0370-2693(91)91013-L}{\emph{Phys. Lett.}
  {\bfseries B272} (1991) 67--70}.

\bibitem{Foot:2014mia}
R.~Foot, \emph{{Mirror dark matter: Cosmology, galaxy structure and direct
  detection}}, \href{https://doi.org/10.1142/S0217751X14300130}{\emph{Int. J.
  Mod. Phys.} {\bfseries A29} (2014) 1430013},
  [\href{https://arxiv.org/abs/1401.3965}{{\ttfamily 1401.3965}}].

\bibitem{Chacko:2005pe}
Z.~Chacko, H.-S. Goh and R.~Harnik, \emph{{The Twin Higgs: Natural electroweak
  breaking from mirror symmetry}},
  \href{https://doi.org/10.1103/PhysRevLett.96.231802}{\emph{Phys. Rev. Lett.}
  {\bfseries 96} (2006) 231802},
  [\href{https://arxiv.org/abs/hep-ph/0506256}{{\ttfamily hep-ph/0506256}}].

\bibitem{Chacko:2005vw}
Z.~Chacko, Y.~Nomura, M.~Papucci and G.~Perez, \emph{{Natural little hierarchy
  from a partially goldstone twin Higgs}},
  \href{https://doi.org/10.1088/1126-6708/2006/01/126}{\emph{JHEP} {\bfseries
  01} (2006) 126}, [\href{https://arxiv.org/abs/hep-ph/0510273}{{\ttfamily
  hep-ph/0510273}}].

\bibitem{Chacko:2005un}
Z.~Chacko, H.-S. Goh and R.~Harnik, \emph{{A Twin Higgs model from left-right
  symmetry}}, \href{https://doi.org/10.1088/1126-6708/2006/01/108}{\emph{JHEP}
  {\bfseries 01} (2006) 108},
  [\href{https://arxiv.org/abs/hep-ph/0512088}{{\ttfamily hep-ph/0512088}}].

\bibitem{Craig:2015pha}
N.~Craig, A.~Katz, M.~Strassler and R.~Sundrum, \emph{{Naturalness in the Dark
  at the LHC}}, \href{https://doi.org/10.1007/JHEP07(2015)105}{\emph{JHEP}
  {\bfseries 07} (2015) 105},
  [\href{https://arxiv.org/abs/1501.05310}{{\ttfamily 1501.05310}}].

\bibitem{Garcia:2015loa}
I.~Garcia~Garcia, R.~Lasenby and J.~March-Russell, \emph{{Twin Higgs WIMP Dark
  Matter}}, \href{https://doi.org/10.1103/PhysRevD.92.055034}{\emph{Phys. Rev.}
  {\bfseries D92} (2015) 055034},
  [\href{https://arxiv.org/abs/1505.07109}{{\ttfamily 1505.07109}}].

\bibitem{Garcia:2015toa}
I.~Garcia~Garcia, R.~Lasenby and J.~March-Russell, \emph{{Twin Higgs Asymmetric
  Dark Matter}},
  \href{https://doi.org/10.1103/PhysRevLett.115.121801}{\emph{Phys. Rev. Lett.}
  {\bfseries 115} (2015) 121801},
  [\href{https://arxiv.org/abs/1505.07410}{{\ttfamily 1505.07410}}].

\bibitem{Craig:2015xla}
N.~Craig and A.~Katz, \emph{{The Fraternal WIMP Miracle}},
  \href{https://doi.org/10.1088/1475-7516/2015/10/054}{\emph{JCAP} {\bfseries
  1510} (2015) 054}, [\href{https://arxiv.org/abs/1505.07113}{{\ttfamily
  1505.07113}}].

\bibitem{Farina:2016ndq}
M.~Farina, A.~Monteux and C.~S. Shin, \emph{{Twin mechanism for baryon and dark
  matter asymmetries}},
  \href{https://doi.org/10.1103/PhysRevD.94.035017}{\emph{Phys. Rev.}
  {\bfseries D94} (2016) 035017},
  [\href{https://arxiv.org/abs/1604.08211}{{\ttfamily 1604.08211}}].

\bibitem{Kusenko:1997zq}
A.~Kusenko, \emph{{Solitons in the supersymmetric extensions of the standard
  model}}, \href{https://doi.org/10.1016/S0370-2693(97)00584-4}{\emph{Phys.
  Lett.} {\bfseries B405} (1997) 108},
  [\href{https://arxiv.org/abs/hep-ph/9704273}{{\ttfamily hep-ph/9704273}}].

\bibitem{Coleman:1977py}
S.~R. Coleman, \emph{{The Fate of the False Vacuum. 1. Semiclassical Theory}},
  \href{https://doi.org/10.1103/PhysRevD.15.2929}{\emph{Phys. Rev.} {\bfseries
  D15} (1977) 2929--2936}. [Erratum:
  \href{https://doi.org/10.1103/PhysRevD.16.1248}{Phys. Rev. D16, 1248
  (1977)}].

\bibitem{Callan:1977pt}
C.~G. Callan, Jr. and S.~R. Coleman, \emph{{The Fate of the False Vacuum. 2.
  First Quantum Corrections}},
  \href{https://doi.org/10.1103/PhysRevD.16.1762}{\emph{Phys. Rev.} {\bfseries
  D16} (1977) 1762--1768}.

\bibitem{Coleman:1977th}
S.~R. Coleman, V.~Glaser and A.~Martin, \emph{{Action Minima Among Solutions to
  a Class of Euclidean Scalar Field Equations}},
  \href{https://doi.org/10.1007/BF01609421}{\emph{Commun. Math. Phys.}
  {\bfseries 58} (1978) 211}.

\bibitem{Linde:1981zj}
A.~D. Linde, \emph{{Decay of the False Vacuum at Finite Temperature}},
  \href{https://doi.org/10.1016/0550-3213(83)90293-6}{\emph{Nucl. Phys.}
  {\bfseries B216} (1983) 421}. [Erratum:
  \href{https://doi.org/10.1016/0550-3213(83)90072-X}{Nucl. Phys. B223, 544
  (1983)}].

\bibitem{ATL-PHYS-PUB-2014-019}
{\scshape ATLAS} collaboration, \emph{{Prospects for measuring Higgs pair
  production in the channel $H(\rightarrow\gamma\gamma)H(\rightarrow
  b\overline{b}) $ using the ATLAS detector at the HL-LHC}}.
\newblock Geneva, Oct, 2014.

\bibitem{ATL-PHYS-PUB-2015-046}
{\scshape ATLAS} collaboration, \emph{{Higgs Pair Production in the
  $H(\rightarrow \tau\tau)H(\rightarrow b\bar{b})$ channel at the
  High-Luminosity LHC}}.
\newblock Geneva, Nov, 2015.

\bibitem{Enqvist:1997si}
K.~Enqvist and J.~McDonald, \emph{{Q balls and baryogenesis in the MSSM}},
  \href{https://doi.org/10.1016/S0370-2693(98)00271-8}{\emph{Phys. Lett.}
  {\bfseries B425} (1998) 309--321},
  [\href{https://arxiv.org/abs/hep-ph/9711514}{{\ttfamily hep-ph/9711514}}].

\bibitem{Frieman:1989bx}
J.~A. Frieman, A.~V. Olinto, M.~Gleiser and C.~Alcock, \emph{{Cosmic Evolution
  of Nontopological Solitons. 1.}},
  \href{https://doi.org/10.1103/PhysRevD.40.3241}{\emph{Phys. Rev.} {\bfseries
  D40} (1989) 3241}.

\bibitem{Griest:1989bq}
K.~Griest and E.~W. Kolb, \emph{{Solitosynthesis: Cosmological Evolution of
  Nontopological Solitons}},
  \href{https://doi.org/10.1103/PhysRevD.40.3231}{\emph{Phys. Rev.} {\bfseries
  D40} (1989) 3231}.

\bibitem{Hardy:2014mqa}
E.~Hardy, R.~Lasenby, J.~March-Russell and S.~M. West, \emph{{Big Bang
  Synthesis of Nuclear Dark Matter}},
  \href{https://doi.org/10.1007/JHEP06(2015)011}{\emph{JHEP} {\bfseries 06}
  (2015) 011}, [\href{https://arxiv.org/abs/1411.3739}{{\ttfamily 1411.3739}}].

\bibitem{Barger:2008jx}
V.~Barger, P.~Langacker, M.~McCaskey, M.~Ramsey-Musolf and G.~Shaughnessy,
  \emph{{Complex Singlet Extension of the Standard Model}},
  \href{https://doi.org/10.1103/PhysRevD.79.015018}{\emph{Phys. Rev.}
  {\bfseries D79} (2009) 015018},
  [\href{https://arxiv.org/abs/0811.0393}{{\ttfamily 0811.0393}}].

\end{thebibliography}\endgroup
\end{document}